\documentclass[aps,prb,twocolumn,showpacs,superscriptaddress]{revtex4}

\usepackage{graphicx}

\begin{document}


\title{Scanning SQUID microscopy of vortex clusters in multiband superconductors}



\author{Taichiro~Nishio}
\affiliation{INPAC-Institute for Nanoscale Physics and Chemistry,
Katholieke Universiteit Leuven, Celestijnenlaan 200D, B-3001
Leuven, Belgium}


\author{Vu~Hung~Dao}
\affiliation{INPAC-Institute for Nanoscale Physics and Chemistry,
Katholieke Universiteit Leuven, Celestijnenlaan 200D, B-3001
Leuven, Belgium}

\author{Qinghua Chen}
\affiliation{INPAC-Institute for Nanoscale Physics and Chemistry,
Katholieke Universiteit Leuven, Celestijnenlaan 200D, B-3001
Leuven, Belgium}

\author{Liviu~F.~Chibotaru}
\affiliation{INPAC-Institute for Nanoscale Physics and Chemistry,
Katholieke Universiteit Leuven, Celestijnenlaan 200D, B-3001
Leuven, Belgium}

\author{Kazuo~Kadowaki}
\affiliation{Institute of Materials Science, University of
Tsukuba, Tsukuba, Ibaraki 305-8573, Japan}

\author{Victor~V.~Moshchalkov}
\email[Corresponding author: victor.moshchalkov@fys.kuleuven.be]{}
\affiliation{INPAC-Institute for Nanoscale Physics and Chemistry,
Katholieke Universiteit Leuven, Celestijnenlaan 200D, B-3001
Leuven, Belgium}


\date{\today}

\begin{abstract}
In type-1.5 superconductors, vortices emerge in clusters, which grow
in size with increasing magnetic field. These novel vortex clusters
and their field dependence are directly visualized by scanning SQUID
microscopy at very low vortex densities in MgB$_{2}$ single
crystals. Our observations are elucidated by simulations based on a
two-gap Ginzburg-Landau theory in the type-1.5 regime.

\end{abstract}

\pacs{74.25.Qt, 74.70.Ad}

\maketitle


In magnesium diboride\cite{nagamatsu}, MgB$_{2}$, the
superconducting gaps open for both the two-dimensional (2D)
$\sigma$-band (gap size $\Delta_{\sigma}$=7.1 meV
\cite{iavarone,roditchev}) and the 3D $\pi$-band ($\Delta_{\pi}$=2.2
meV\cite{roditchev,eskildsen}). The possibility of type-1.5
superconductivity\cite{moshchalkov} has been suggested for clean
single crystals of MgB$_{2}$, which lie in the type-1.5 regime:
$\kappa_{\sigma}$=3.7$>1/\sqrt{2}$ (type-2) and
$\kappa_{\pi}$=0.66$<1/\sqrt{2}$ (type-1), where $\kappa_{\alpha}$
($\alpha=\sigma,\pi$) is the Ginzburg-Landau (GL) parameter for each
band estimated from the band structure calculations\cite{mazin}. In
contrast to conventional type-2 superconductors\cite{brandt}, in
type-1.5 superconductors, vortex stripes and gossamer-like vortex
patterns, i.e., vortex clusters emerge at relatively low applied
fields,\cite{moshchalkov} which is due to a competition between
attractive (type-1) and repulsive (type-2) vortex interactions
governed by a two-gap GL
theory\cite{askerzade,babaev1,babaev2,babaev3,dao}. Interestingly, a
substantial difference in vortex structure between type-1.5 and
type-2 superconductors, as a fingerprint of type-1.5
superconductivity, is expected at very low vortex densities because
the intervortex distances in clusters are likely to be almost
independent of the applied field in type-1.5 superconductors,
whereas the intervortex distances in type-2 superconductors follow
the conventional dependence $(\phi_{0}/B)^{1/2}$, where
$\phi_{0}$(=2.07$\times$$10^{-15}$ T m) is the flux quantum and $B$
the magnetic field. This motivates strongly direct vortex
visualization experiments aimed at investigating vortex structure at
very low vortex densities in MgB$_{2}$ single crystals.
Additionally, it is important to verify the existence of type-1.5
superconductivity in high quality MgB$_{2}$ crystals different from
the ones used in Ref. \onlinecite{moshchalkov}.


In order to study vortex structure in MgB$_{2}$, we made scans
with a scanning superconducting quantum interference device
(SQUID) microscope on single crystals of MgB$_{2}$ which were
grown by a pressure synthesis technique\cite{machida} (a crystal
grown by this technique was used elsewhere\cite{souma}). The
superconducting transition temperatures $T_{c}$ of the crystals
measured by a SQUID magnetometer are 38.5 K [the transition width
$\Delta$$T_{c}$=0.8 K (10-90 \% criterion)]. The surface of
crystals was confirmed not to have any cracks or holes by using a
field emission scanning electron microscope in the secondary
electron regime. High-resolution transmission electron microscope
images and electron diffraction patterns for the crystals show
that the crystals have no grain boundaries. No impurity
contamination was detected within 0.1$\%$ accuracy by an electron
probe microanalyzer.

Our scanning SQUID microscope is based on prototype
SQM2000\cite{nishio} (SII NanoTechnology Inc.). We fabricated a
SQUID sensor, which is a SQUID magnetometer linked with a circular
Nb pickup loop. The Nb pickup loop with an inner diameter of 8
$\mu$m scans the surface of a sample, keeping a few micrometers
away from the surface. A magnetic flux sensitivity is
5~$\mu\phi_{0}$/Hz$^{1/2}$. A field noise referred to the pickup
loop is $\sim$10$^{-10}$ T/Hz$^{1/2}$. A spatial resolution is
$\sim$2.5 $\mu$m. Before scans, as-grown crystals were cooled from
a temperature higher than $T_{c}$ to 4.2 K under the magnetic
field parallel to the $c$-axis and then scans were made on the
$ab$-plane at 4.2 K. Scanning steps are 0.5 $\mu$m and 1.0 $\mu$m
for MgB$_{2}$ single crystals and a Nb film, respectively.

\begin{figure}
\includegraphics[width=0.8\linewidth]{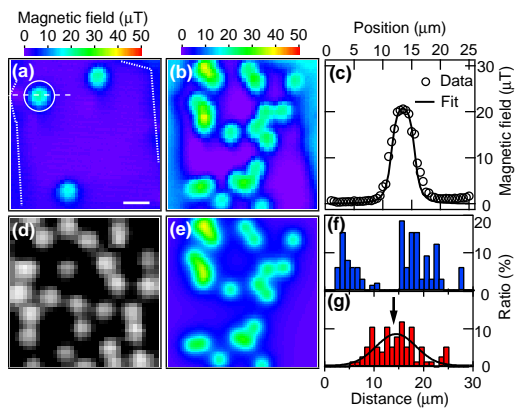}
\caption{\label{Fig. 1} (color online) Scanning SQUID microscope
images of vortices at (a) 1 $\mu$T and (b) 10 $\mu$T for a
MgB$_{2}$ single crystal, the dimensions of which are
approximately 80$\times$60 $\mu$m$^{2}$. Dotted lines show the
edges of the sample. A scale bar corresponds to 10 $\mu$m. The
integral of the magnetic field within a solid line in (a) gives
the flux $\sim$$\phi_{0}$. (c) The cross section of the vortex
image along a broken line in (a) (open circles) and the model fit
(solid line)(see text). (d) Scanning SQUID microscope image of
vortices at 10 $\mu$T in a Nb film with a thickness of 200 nm
($T_{c}$=9.1 K). (e) The 2D multi-fits of the SQUID image in (b)
to numerical calculations by a London model (see text). (f) and
(g) The distribution of the intervortex distance derived from (e)
and (d), respectively. A solid line in (g) represents a fit to the
Gaussian function. An arrow indicates the intervortex distance
calculated from $(\phi_{0}/B)^{1/2}$.}
\end{figure}


We model vortex systems in a type-1.5 superconductor in the same
way as in Ref. \onlinecite{moshchalkov}. We invoke the following
two-gap GL theory to derive the vortex-vortex interaction
numerically. A two-gap GL free energy
functional\cite{askerzade,babaev1,babaev2,babaev3,dao} is the sum
of two single-band GL functionals with the Josephson coupling term
corresponding to the interaction between two bands:
\begin{equation}
    F=\int dr^{3}(F_{\sigma}+F_{\pi}+F_{\sigma\pi}+\frac{h^{2}}{2\mu_{0}}),
\end{equation}
where $F_{\alpha}$ ($\alpha$=$\sigma,\pi$) is the free energy of
each band and $h$ (=$|\textbf{h}|$) is a magnetic field
(\textbf{h}=~rot~$\textbf{A}$, where $\textbf{A}$ is the vector
potential):\cite{tinkham}
\begin{equation}
    F_{\alpha}=2E_{\alpha}|\Psi_{\alpha}|^{2}+|E_{\alpha}||\Psi_{\alpha}|^{4}+C|(-i\nabla+\frac{2\pi}{\phi_{0}}\textbf{A})\Psi_{\alpha}|^{2},
\end{equation}
where $E_{\alpha}$ is the condensation energy\cite{eisterer} and
$C$=$(\phi_{0}/2\pi)\sqrt{|E_{\alpha}|/\mu_{0}\kappa_{\alpha}^{3}}$.
$F_{\sigma\pi}$ is the Josephson coupling term:
$F_{\sigma\pi}=\frac{E_{\gamma}}{2}(\Psi_{\sigma}^{\ast}\Psi_{\pi}+\Psi_{\pi}^{\ast}\Psi_{\sigma})$,
where $E_{\gamma}$ is the coupling energy, where we take a value
used in Ref. \onlinecite{eisterer}. The vortex-vortex interaction
energy between two vortices is the sum of increased energies of
each vortex caused by the presence of the other. It can be derived
by numerically minimizing the two-gap GL free energy of two
vortices with a variational procedure \cite{jacobs}. The
minimization gives the interaction consisting of short-range
repulsion and weak long-range attraction, which is similar to that
reported in Ref. \onlinecite{jacobs}.

\begin{figure}
\includegraphics[width=0.8\linewidth]{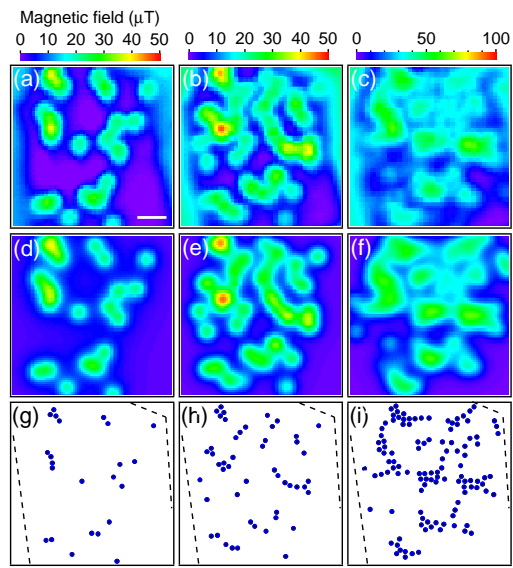}
\caption{\label{Fig. 2} (color online) Scanning SQUID microscope
images of vortices at (a) 10 $\mu$T, (b) 20 $\mu$T, and (c) 50
$\mu$T for a MgB$_{2}$ single crystal. A scale bar corresponds to
10 $\mu$m. (d)-(f) The 2D multi-fits of the SQUID images in
(a)-(c) by a London model (see text). (g)-(i) The locations of
vortices in (d)-(f) obtained by the 2D multi-fits of the images.
Broken lines show the edges of the sample.}
\end{figure}

An equation of motion for a vortex $\emph{i}$ is
$\textbf{F}_{i}=\textbf{F}_{i}^{vv}+\textbf{F}_{i}^{T}=\eta
\textbf{v}_{i}$, where $\textbf{F}_{i}^{vv}$ is the vortex-vortex
interaction term we calculate above, $\textbf{F}_{i}^{T}$ the
thermal stochastic force which satisfies $\langle
\textbf{F}_{i}^{T}(t)\rangle$=0 and
$\langle\textbf{F}_{i}^{T}(t)\textbf{F}_{j}^{T}(t')\rangle=2\eta\delta_{ij}\delta(t-t'
)k_{B}T$, and $\eta$ the viscosity, where we take a value
estimated in Ref. \onlinecite{moshchalkov}. The systems with no
pinning centers were initially prepared in a high temperature
molten state and then annealed with two million temperature
steps.\cite{reichhardt} We made the system stable during 2000 time
steps in each step of temperature.

Figures 1(a) and 1(b) show a scanning SQUID microscope images of
vortices at very low vortex densities for a MgB$_{2}$ single
crystal. The size of a vortex in the image does not correspond
straightforwardly to the penetration depth $\lambda$ but rather to a
combination of $\lambda$ and the extent of a stray field emanating
from a vortex at the sample surface. In order to estimate $\lambda$
it is necessary to compare the cross section of the vortex image
with appropriate numerical calculations where the extent of the
stray field is taken into account.\cite{nishio} We calculate
numerically the spatial distribution of a magnetic field from a
vortex above the sample surface, using a London
model\cite{nishio,kogan,kirtley}.

Figure 1(c) shows a fit of the cross section of a vortex image in
Fig. 1(a) by the numerical simulations. We obtain
$\lambda$=0.01-0.13 $\mu$m ($z$=1.7$\pm$0.1 $\mu$m, where $z$ is
the distance between the sample surface and the pickup loop) from
the fit. The $z$ value was determined by the fit of a vortex image
in a Nb film (known parameter $\lambda$=50 nm) mounted on
 a sample holder with a MgB$_{2}$ crystal. The range of $\lambda$
 is a residual based on the error bar in the $z$ value which was
 assigned by doubling of the $\chi$ value in the fit in a Nb film
 (a doubling of the $\chi$ value only gives statistical errors).
 Due to the error it is difficult to specify how close our
 $\lambda$ value is to the band structure calculations\cite{mazin}.
 However, first-principles calculations\cite{mazin,golubov} of the
penetration depth in the clean limit fall on our $\lambda$ value
within the error bar.

In type-2 superconductors with the weak pinning, the Abrikosov
lattice is often seen in field cooling experiments even if a vortex
density is quite low:\cite{nishio} the vortex lattice made of
vortices strongly interacting with each other at temperatures very
close to $T_{c} $ is frozen with decreasing temperature. Vortex
patterns at very low vortex densities in a MgB$_{2}$ single crystal,
which has the weak pinning,\cite{eisterer} seem to be different from
that in the conventional weak pinning system. At $B$=1 $\mu$T
vortices are located far away from each other.[Fig. 1(a)] However,
with increasing applied field vortices come closer to each other to
form compact groups as shown in Figs. 1(b) and 1(e), which is never
seen in superconductors with the weak pinning, such as MoGe
films\cite{nishio} and NbSe$_{2}$ single crystals\cite{moshchalkov}.

\begin{figure}
\includegraphics[width=0.8\linewidth]{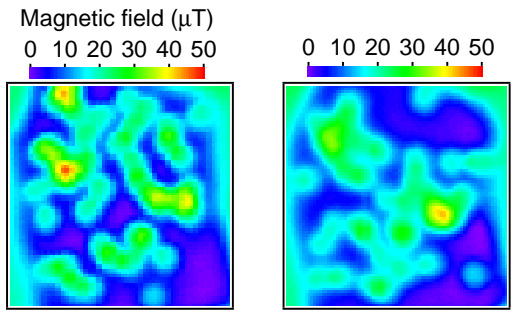}
\caption{\label{Fig. 3} Scanning SQUID microscope images of
vortices at 20 $\mu$T obtained from different cool downs. The left
image is the same as Fig. 2(b).}
\end{figure}

In a Nb film with the strong pinning, disordered vortex patterns are
observed at low vortex densities, as shown in Fig. 1(d). We now
compare qualitatively these vortex patterns. Figure 1(e) shows the
2D multi-fits\cite{nishio2} of the scanning SQUID image in Fig. 1(b)
to numerical calculations from a London model\cite{kogan, kirtley}.
Fits were made by parameterizing the coordinates of each position of
vortices with $z$ and $\lambda$ determined above. From the fits it
turns out that the number of vortices is 23, which is indeed
consistent with 23 estimated by $SB/\phi_{0}$, where $S$ is the
areas of a sample. The distribution of the intervortex distance can
be derived from the fits, as shown in Fig. 1(f). This distribution
is a bimodal and it clearly splits into two: one ranging from
$\sim$2 $\mu$m to $\sim$7 $\mu$m and another from $\sim$15 $\mu$m to
$\sim$30 $\mu$m. In principle, the former corresponds to the
intragroup distribution and the latter corresponds to that
concerning the intergroup. On the other hand, the vortex
distribution in a Nb film can be fitted by the Gaussian function, as
shown in Fig. 1(g). The center of the peak in the Gaussian function
is at 14.5 $\mu$m, indicating that the average intervortex distance
is 14.5 $\mu$m. This agrees with the intervortex distance
$(\phi_{0}/B)^{1/2}$ of $\sim$14 $\mu$m estimated by assuming the
presence of the Abrikosov lattice. This fact indicates that the
vortex pattern in Fig. 1(b) is not a disordered Abrikosov lattice
but vortex cluster array. On the other hand, the vortex pattern in
Fig. 1(d) does correspond to the strongly disordered Abrikosov
lattice.

Figures 2(a)-2(c) show the variation of the vortex structure with
applied fields. After vortices form local clusters in Fig. 2(a),
some islands of magnetic flux and the semi-Meissner regions come out
of the grouping with increasing applied field. [Figs. 2(b) and 2(c)]
The positions of the vortices can be determined by the 2D multi-fits
[Figs. 2(d)-2(f)] of Figs 2(a)-2(c), as shown in Figs. 2(g)-2(i).
The position maps show that vortices line up locally in the early
stage [Fig. 2(g)] and then small vortex clusters merge with each
other, [Fig. 2(h)] to grow into bigger clusters with applied fields.
[Fig. 2(i)] The number of vortices changes to 49 in Fig. 2(h) and
then 114 in Fig. 2(i). These are reasonably well approximated by 46
and 116 estimated from $SB/\phi_{0}$, respectively.

\begin{figure}
\includegraphics[width=1.0\linewidth]{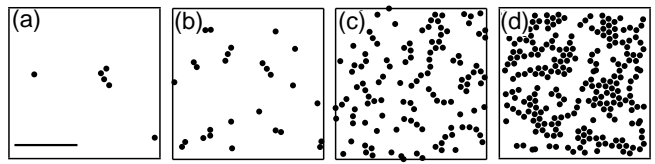}
\caption{\label{Fig. 4} Simulations of systems consisting of (a) 6,
(b) 27, (c) 122, and (d) 500 vortices which have the interaction
energy calculated from a two-gap GL theory for the type-1.5
condition (see text). A scale bar corresponds to 10 $\mu$m.}
\end{figure}

We note that the vortex array varies considerably in the locations
of clusters measured by different cool downs, while keeping the
same general features, as seen in Fig 3. Since the pinning sites
are fixed, this implies that the vortex clustering is not due to
the presence of pinning centers but is rather to be attributed to
the peculiar vortex-vortex interaction combining repulsion and
attraction.

Figures 4(a)-4(d) show results of the simulation mentioned above. In
type-1.5 superconductors vortices begin to form clusters consisting
of a few vortices at very low vortex densities, [Fig. 4(a)] which
does not coincide with the experiment displayed in Fig. 1(a).
Perhaps, the existence of some pinning centers may be responsible
for this fact. This will be considered in the future work. The
number of small clusters increases with applied fields, [Fig. 4(b)]
which apparently corresponds to Fig. 2(g). Then small clusters start
to merge with each other,[Fig. 4(c)] to grow into bigger clusters
with a further increase of an applied field [Fig. 4(d)]. This
simulation result coincides quite well with experimental data [Figs.
2(h) and 2(i)].

The simulations show that in a vortex cluster the distances between
vortices and their nearest neighbors are almost independent of the
applied field ($\sim$1 $\mu$m). Therefore, these distances are not
determined by the expression $(\phi_{0}/B)^{1/2}$ but rather by the
competition between short-range repulsive and long-range attractive
interactions between vortices. In Fig. 5 the average intervortex
distance in clusters obtained from scanning SQUID and decoration
images\cite{moshchalkov} is shown as a function of applied field,
compared with the dependence $(\phi_{0}/B)^{1/2}$. The average
distances indeed do not follow any dependence close to
$(\phi_{0}/B)^{1/2}$ but follow instead a very weak linear field
dependence. The intervortex distance in clusters, independent of the
applied field, can be identified as a unique property of a system of
vortices with short-range repulsion and long-range attraction, i.e.,
vortices induced in a type-1.5 superconductor. This fact indicates
that vortex clusters definitely emerge even in crystals grown by a
different crystal maker, implying that in general vortices tend to
form vortex clusters in high quality single crystals of
MgB$_{2}$.\cite{moshchalkov}

\begin{figure}
\includegraphics[width=0.8\linewidth]{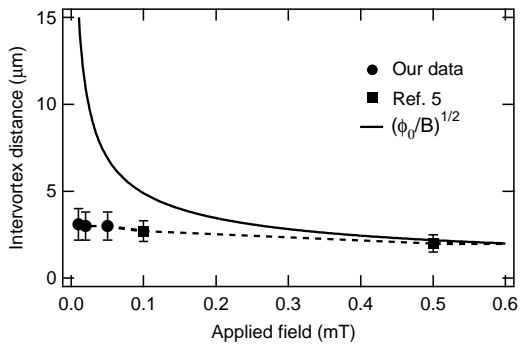}
\caption{\label{Fig. 5} The average intervortex distance in vortex
clusters as a function of applied field. These distances are
obtained from fits of scanning SQUID images in Figs. 2(g)-2(i)
(filled circles) and decoration images\cite{moshchalkov} at 0.1 mT
and 0.5 mT (filled squares). Error bars indicate the range of the
distribution of vortex distances in clusters. A solid line
represents the dependence $(\phi_{0}/B)^{1/2}$.}
\end{figure}

A difference between $(\phi_{0}/B)^{1/2}$ and the intervortex
distance in clusters decreases as an applied field increase and
these two length scales become equivalent around 0.6 mT. Any
vortex cluster is unlikely to be observed above 0.6 mT. In fact
Moshchalkov et al.\cite{moshchalkov} have reported no clearly
visible vortex clusters at 1.0 mT. The observations of long-range
Abrikosov lattices at 20 mT and 0.2 T\cite{vinnikov,eskildsen} are
consistent with this scenario. Although Vinnikov et
at.\cite{vinnikov} have obtained a decoration image of a
disordered vortex lattice at 0.44 mT, it would be too difficult to
find vortex clusters, a size of which is more than 20 $\mu$m
around 0.5 mT, as reported in Ref. \onlinecite{moshchalkov}.


In conclusion, scanning SQUID microscopy shows that in high quality
MgB$_{2}$ single crystals vortices line up locally at very low
vortex densities and grow into vortex clusters with increasing
applied field. The average distances between nearest neighboring
vortices forming clusters are practically independent of the applied
field $B$, in contrast to the conventional (1/$B)^{1/2}$ behavior.
The simulations of a vortex system of a type-1.5 superconductor
indeed reproduce these observations. Our experimental data, as well
as consistency between experiment and the simulations, clearly
demonstrates that high quality MgB$_{2}$ single crystals lie in the
type-1.5 regime.

This work was supported by Methusalem Funding by the Flemish
government, the Flemish FWO, the Belgian Interuniversity
Attraction Poles (IAP), the ESF-NES ``Nanoscience and Engineering
in Superconductivity" Programmes, and JSPS core-to-core NES
program.

\newpage


\begin{thebibliography}{}



\bibitem{nagamatsu} J. Nagamatsu, N. Nakagawa, T. Muranaka, Y. Zenitani, and J. Akimitsu  Nature {\bf 410}, 63
(2001).

\bibitem{iavarone} M. Iavarone, G. Karapetrov, A. E. Koshelev, W. K. Kwok, G. W. Crabtree, D. G. Hinks,
W. N. Kang, E. M. Choi, H. J. Kim, H. J. Kim, and S. I. Lee, Phys.
Rev. Lett. {\bf 89}, 187002 (2002); P. Szab\'{o}, P. Samuely, J.
Ka\v{c}mar\v{c}\'{i}k, T. Klein, J. Marcus, D. Fruchart, S.
Miraglia, C. Marcenat, and A. G. M. Jansen, \emph{ibid}
\textbf{87}, 137005 (2001).

\bibitem{roditchev} F. Giubileo, D. Roditchev, W. Sacks, R. Lamy, D. X. Thanh, J. Klein, S. Miraglia,
 D. Fruchart, J. Marcus, and P. Monod, Phys. Rev. Lett. \textbf{87}, 177008 (2001).

\bibitem{eskildsen} M. R. Eskildsen, M. Kugler, S. Tanaka, J. Jun, S. M. Kazakov, J. Karpinski,
 and {\O}. Fischer, Phys. Rev. Lett. {\bf89}, 187003 (2002); G. Rubio-Bollinger, H. Suderow, and S. Vieira
\emph{ibid} \textbf{86}, 5582 (2001).

\bibitem{moshchalkov} V. V. Moshchalkov, M. Menghini, T. Nishio, Q. H. Chen, A. V. Silhanek, V. H. Dao,
 L. F. Chibotaru, N. D. Zhigadlo, and J. Karpinski, Phys. Rev. Lett.
\textbf{102}, 117001 (2009).

\bibitem{mazin} I. I. Mazin, O. K. Andersen, O. Jepsen, O. V. Dolgov, J. Kortus, A. A. Golubov,
A. B. Kuz'menko, and D. van der Marel, Phys. Rev. Lett. {\bf 89},
107002 (2002).

\bibitem{brandt} E. H. Brandt, Rep. Prog. Phys. \textbf{58}, 1465
(1995).

\bibitem{askerzade} I. N. Askerzade, A. Gencer, and N. G\"{u}\c{c}l\"{u},
Supercond. Sci. Technol. {\bf 15}, L17 (2002).

\bibitem{babaev1} E. Babaev, Phys. Rev. Lett. {\bf 89}, 067001
(2002).

\bibitem{babaev2} E. Babaev and M. Speight, Phys. Rev. B {\bf 72},
180502(R) (2005).

\bibitem{babaev3} E. Babaev and N. W. Ashcroft, Nature Physics {\bf
3}, 530 (2007).

\bibitem{dao} M. E. Zhitomirsky and V. H. Dao, Phys. Rev. B {\bf
69}, 054508 (2004).

\bibitem{machida} Y. Machida, S. Sasaki, H. Fujii, M. Furuyama, I. Kakeya, and K. Kadowaki
, Phys. Rev. B {\bf 67}, 094507 (2003).

\bibitem{souma} S. Souma, Y. Machida, T. Sato, T. Takahashi, H. Matsui, S. C. Wang,
 H. Ding, A. Kaminski, J. C. Campuzano, S. Sasaki, and K. Kadowaki, Nature \textbf{423}, 65 (2003).

\bibitem{nishio} T. Nishio, S. Okayasu, J. I. Suzuki, N. Kokubo, and
K. Kadowaki, Phys. Rev. B {\bf 77}, 052503 (2008).

\bibitem{tinkham} M. Tinkham, \emph{Introduction to Superconductivity},
2nd ed. (McGraw-Hill, New York, 1996).

\bibitem{eisterer} M. Eisterer, Supercond. Sci. Technol.
\textbf{20}, R47 (2007).

\bibitem{jacobs} L. Jacobs and C. Rebbi, Phys. Rev. B \textbf{19},
4486 (1979).

\bibitem{reichhardt} C. Reichhardt, C. T. Olson, I. Martin, and A.
R. Bishop, Europhys. Lett. \textbf{61}, 221 (2003).



\bibitem{kogan} V. G. Kogan, V. V. Dobrovitski, J. R. Clem, Y.
Mawatari, and R. G. Mints, Phys. Rev. B \textbf{63}, 144501
(2001).

\bibitem{kirtley} J. R. Kirtley, V. G. Kogan, J. R. Clem, and K.
A. Moler, Phys. Rev. B \textbf{59}, 4343 (1999).


\bibitem{golubov} A. A. Golubov, A. Brinkman, O. V. Dolgov, J.
Kortus, and O. Jepsen, Phys. Rev. B \textbf{66}, 054524 (2002).

\bibitem{nishio2} T. Nishio, Q. Chen, W. Gillijns, K. De Keyser, K. Vervaeke, and V. V. Moshchalkov,
 Phys. Rev. B \textbf{77},
012502 (2008).

\bibitem{vinnikov} L. Y. Vinnikov, J. Karpinski, S. M. Kazakov, J. Jun,
J. Anderegg, S. L. Bud'ko, and P. C. Canfield, Phys. Rev. B {\bf
67}, 092512 (2003).

\end{thebibliography}
\end{document}